\documentclass{aa} 
\usepackage{natbib}
\usepackage{graphicx}
\usepackage{txfonts}
\usepackage{epsfig}
\usepackage{times}
\usepackage{rotating}
\usepackage{float}
\usepackage[caption = false]{subfig}
\usepackage{setspace}
\usepackage{lscape}
\usepackage{epstopdf}
\usepackage{tabularx}
\usepackage{caption}
\usepackage[usenames]{color}
\usepackage[toc,page]{appendix}
\usepackage{soul}
\usepackage{amsmath}

\newcommand{\Msun}{M_\odot}

\begin{document}

\title{From relics to stripped systems: the environmental origin of compact galaxies}
\author{Guangwen Chen\inst{1,2}, J. Alfonso L. Aguerri\inst{1,2}\fnmsep\thanks{Corresponding author: jalfonso@iac.es}, Carlos del Burgo\inst{1,2}, Stefano Zarattini\inst{3}}


\institute{Instituto de Astrof\'isica de Canarias, calle Vía L\'actea s/n, E-38205 La Laguna, Tenerife, Spain 
\and Departamento de Astrof\'isica, Universidad de La Laguna, E-38206 La Laguna, Tenerife, Spain
\and Centro de Estudios de F\'isica del Cosmos de Arag\'on (CEFCA), Plaza San Juan 1, 44001 Teruel, Spain}

\date{\today}

\abstract
{Compact galaxies represent a key population for understanding galaxy evolution, as their high stellar densities are closely linked to early formation and/or stripping processes. However, the relative role of internal structure and environment in shaping their origin and evolution remains unclear.}
{We aim to investigate how the abundance and quenching of compact galaxies depend on environment, and to determine whether their formation is governed by a single evolutionary pathway or multiple channels.
}
{We analyse a large sample of galaxies in the nearby Universe ($z < 0.05$) with stellar masses in the range $8.2 < \log(M_{\star}/M_{\odot})<10.5$, comparing compact and control populations as a function of local overdensity by measuring the fraction of galaxies above a given overdensity ($\delta$) threshold.}
{We find that the environmental dependence of compact galaxies is non-monotonic, with two characteristic overdensities, the transition overdensity ($\delta_{\rm t}$) and the critical overdensity ($\delta_{\rm c}$), defining three distinct regimes. Below $\delta_{\rm t}$, compact galaxies are relatively more abundant than the control population, while between $\delta_{\rm t}$ and $\delta_{\rm c}$ their relative abundance decreases significantly. Above $\delta_{\rm c}$, compact galaxies become increasingly overabundant again, although this behaviour strongly depends on stellar mass. Low-mass ($8.2 < \log(M/M_{\odot}) < 9.0$) compact galaxies exhibit the strongest deficit at intermediate overdensities. In contrast, high-mass compact galaxies ($9.0 < \log(M/M_{\odot}) < 10.5$) dominate the excess population above $\delta_{\rm c}$. In addition, we find that the fraction of red galaxies increases with local overdensity in control and compact samples. However, the red galaxy fraction is systematically higher in compact galaxies than in the control sample, indicating that both environment and internal structure (compactness) play a role in galaxy quenching. The relative importance of these effects strongly depends on stellar mass, with environmental quenching being more prominent in low-mass compact galaxies, while internal structural properties may dominate in the high-mass compact population.}
{Our results support a unified evolutionary framework in which compact galaxies do not arise from a single evolutionary channel, but instead reflect multiple pathways whose relative importance depends strongly on stellar mass and environment. Low-mass compact galaxies appear to be closely linked to environmentally driven transformation and quenching processes acting in groups, filaments, and cluster infall regions. In contrast, high-mass compact galaxies are found across all environments, but become increasingly overabundant relative to the control population in the densest regions. Their uniformly high red galaxy fractions and stellar surface mass densities indicate that they survive efficiently in high-density environments as structurally dense systems, similar to compact relic galaxies.}


\authorrunning{Chen et al} 
\titlerunning{the environmental origin of compact galaxies} 
\maketitle

\section{Introduction}
\label{sec:intro}

Compact galaxies, characterised by their small sizes and high stellar mass densities, occupy the extreme tail of the galaxy size--mass relation in the local Universe, with effective radii $\lesssim 1-2\mathrm{kpc}$ and stellar masses $\gtrsim\,10^8\Msun$, and encompass a heterogeneous set of systems including compact ellipticals, blue compact dwarfs and compact relics \citep[e.g.,][]{Faber1973,Kunth2000,Chilingarian2009,Ferre-Mateu2017,Deeley2023,GrebolTomas2023}.   As such, they serve as valuable laboratories for studying the physical processes that drive galaxy transformation, particularly the interplay between internal and environmental processes such as dissipative compaction, mergers, and tidal stripping \citep[e.g.,][]{vanderWel2014,Tacchella2016,Naab2017}. Understanding the formation and evolution of compact galaxies is therefore key for constraining how galaxies assemble and evolve structurally across cosmic time. 

However, despite extensive observational and theoretical efforts, the origin of compact galaxies and the processes responsible for their structural evolution remain uncertain, with several formation scenarios proposed in the literature. Historically, compact galaxies have usually been interpreted as products of environmental transformation processes. In particular, compact galaxies are frequently observed in the vicinity of massive galaxies, where tidal forces can remove the outer stellar components of galaxies, leaving behind dense and metal-rich remnants \citep[e.g.,][]{Chilingarian2009,Huxor2011,Norris2014,Janz2016,Ferre-Mateu2018}. A classical example is the compact elliptical galaxy M32  \citep[e.g.,][]{delBurgo2001,Bekki2001,Graham2002,Choi2002,DSouza2018,Escala2025}, widely interpreted as the tidally stripped remnant of a more massive progenitor orbiting M31.

Compact galaxies are also commonly observed in denser environments such as galaxy groups and clusters, where environmental transformation processes are expected to be particularly efficient \citep[e.g.,][]{Chilingarian2007,Chilingarian2009,Ferre-Mateu2018,Sharonova2025}. In these environments, gravitational mechanisms such as tidal stripping, harassment, and mergers can strongly alter galaxy morphology and remove substantial fractions of stellar mass \citep[e.g.,][]{Hernquist1992,Moore1996,Aguerri2009,Chen2018,Zhang2020,Guo2022}, while hydrodynamical effects such as ram-pressure stripping can efficiently remove gas and, in some cases, induce central star formation episodes through gas compression \citep[e.g.,][]{Gunn1972,Quilis2000,Boselli2006}. These processes may either reduce galaxy sizes by stripping their outer regions \citep[e.g.,][]{Aguerri2004,Huxor2011,Ferre-Mateu2021} or increase central stellar densities through dissipative starbursts \citep[e.g.,][]{Mihos1996,Du2019,Zhang2020b,Bian2025}. 

However, compact galaxies are not exclusively products of environmental transformations. In lower-density environments, some compact galaxies may form as intrinsically compact systems and subsequently experience little structural evolution \citep[e.g.,][]{Huxor2013,Rey2021,Deeley2023,Yi2025}. Merger-driven dissipative compaction has also been proposed for isolated compact early-type galaxies \citep[e.g.,][]{Paudel2014,Ferre-Mateu2018,GrebolTomas2023}. Examples of local massive relic galaxies include systems such as MRK~1216 and PGC~032873, which are widely interpreted as survivors of the high-redshift compact galaxy population (``red nuggets'') that have undergone little size evolution since their formation \citep[e.g.,][]{vanDokkum2010, Ferre-Mateu2017}. These objects provide direct evidence that compact galaxies can persist in isolation over cosmic time, supporting a formation scenario driven by early assembly followed by passive evolution. Nevertheless, relic galaxies are preferentially found today in dense environments rather than in isolation. High-density environments such as galaxy clusters may suppress major mergers and therefore could limit subsequent size growth of galaxies \citep[e.g.,][]{Ostriker1980,Davison2020,Benavides2020,Sureshkumar2024}. The deep gravitational potentials of compact systems may make them more resistant to tidal disruption than extended galaxies. This suggests that dense environments may not only produce compact galaxies through stripping processes, but may also preserve compact systems formed at early epochs by preventing their later structural evolution \citep[e.g.,][]{Ferre-Mateu2018,Ferre-Mateu2021,Carr2024,Caso2024}. This could be the case for NGC 1277 in the Perseus cluster \citep[][]{Trujillo2014,Ferre-Mateu2017}, as well as for a relic-like compact elliptical galaxy, FS90~192, in the Antlia cluster \citep[][]{Caso2024}.

In recent years, wide-field imaging and spectroscopic surveys have enabled the construction of increasingly large samples of compact galaxies, with sample sizes ranging from dozens to nearly 200 objects \citep[e.g.,][]{Chilingarian2015,Tortora2018,Scognamiglio2020,Kim2020,Chen2022,GrebolTomas2023}. These studies have shown that compact galaxies span a broad range of stellar masses, sizes, and structural properties, and appear to form a continuous sequence with other compact stellar systems such as ultra-compact dwarfs, as well as with dwarf ellipticals \citep[e.g.,][]{Norris2014,Zhang2015,Liu2015,Janz2016,Wang2023}. The availability of larger samples has also enabled the first statistical studies of compact galaxies across different environments, revealing a complex and non-uniform environmental dependence.
 
However, current samples remain relatively small and heterogeneous, and a clear picture linking compactness, environment, stellar mass, and star formation activity is still lacking. In addition, many studies on compact galaxies are biased towards relatively massive systems \citep[e.g.,][]{Tortora2018,Scognamiglio2020,GrebolTomas2023} and cannot provide complete samples that extend to low stellar masses ($M_\star \lesssim 10^9\,M_\odot$), where compact galaxies are also known to exist. This limits our ability to explore the full diversity of compact systems and to connect low-mass compact galaxies with their higher-mass counterparts within a unified framework. Particularly, systematic investigations of how the abundance and properties of compact galaxies vary with local environment remain scarce. 

Consequently, it is still unclear how the various proposed formation pathways are connected, and whether their relative importance depends on galaxy properties such as stellar mass and structure, and/or on environment. In particular, it remains an open question whether compact galaxies represent a heterogeneous population arising from multiple evolutionary channels, or a more uniform class shaped by a dominant formation mechanism \citep[e.g.,][]{Norris2014,Janz2016,Kim2020,Ferre-Mateu2021,Deeley2023,Caso2024}. A unified environmental and mass-complete perspective is therefore required to connect the different formation scenarios to specific physical conditions and to disentangle the relative roles of internal processes (nature) and environmental effects (nurture) in shaping compact galaxies.

With the advent of wide-area imaging and spectroscopic surveys such as the Sloan Digital Sky Survey Data Release 16 \citep[SDSS-DR16,][]{Ahumada2020} and Dark Energy Spectroscopic Instrument Data Release 1 \citep[DESI-DR1,][]{DESI2026}, it is now possible to construct statistically significant samples of compact galaxies in the local Universe with well-characterised environments and broad stellar mass coverage. In this work, we investigate the environmental dependence of compact galaxies by analysing their distribution as a function of local overdensity. By adopting a continuous measure of environment, we explore how the relative abundance of compact systems varies across a wide range of environments in a uniform and quantitative manner. Additionally, we examine how their structural properties and quenching behaviour depend on both stellar surface mass density and environment. This approach enables a more general and self-consistent assessment of the connection between galaxy structure, star formation, and environment, and allows us to test whether the environmental dependence of compact galaxies reflects smooth trends or transitions between distinct physical regimes. 

This paper is organised as follows. In Section \ref{sec:data}, we describe the data and the construction of the compact galaxy sample, as well as the definition of the control sample and the environmental characterisation. The main results are presented in Sect. \ref{sec:result}. Sections \ref{sec:discussion} and \ref{sec:conclusions} show the discussion and conclusions of the paper.  Throughout this work, we adopt a $\Lambda$CDM cosmology with $\Omega_m = 0.3$, $\Omega_\Lambda = 0.7$, and $H_0 = 70\,\mathrm{km\,s^{-1}\,Mpc^{-1}}$.

\section{The compact galaxy catalogue} \label{sec:data}

The goal of this section is to construct a homogeneous, statistically significant, and well-defined catalogue of compact galaxies in the local Universe, suitable for studying their dependence on the local environment.

\subsection{Data and sample definition}

We have used the galaxy catalogue from the SDSS-DR16 \citep[][]{Ahumada2020}, covering a large sky area within $130<\mathrm{RA(J2000)}<230$ and $0<\mathrm{DEC(J2000)}<60$ \citep[see][hereafter ZA25]{Zarattini2025}. In particular, we downloaded all objects classified as galaxies in that catalogue. In addition to the photometric parameters of the objects (g- and r-band magnitudes, and the radii containing 50\% and 90\% of the light, $R_{50}$ and $R_{90}$, respectively, we also retrieved the redshifts of the galaxies.  This galaxy catalogue has already been used to study galaxy properties, such as star formation, across different large-scale environments (see ZA25). In addition, it has been used to derive galaxy density profiles of large-scale filaments up to $z < 0.3$ \citep[see][]{Aguerri2026}.

The spectroscopic completeness of the SDSS-DR16 main galaxy sample is limited to an apparent magnitude of $m_r = 17.77$, which corresponds to galaxies with absolute magnitudes of $M_r \sim -19.0$ at $z = 0.05$, or typical stellar masses of $M_\star \sim 10^{9.5}\,M_\odot$. As a result, the SDSS sample alone is incomplete at lower stellar masses, limiting our ability to explore the population of compact galaxies to galaxies with stellar masses smaller than $10^{9}\,M_{\odot}$. To improve the spectroscopic completeness and extend the sample towards lower stellar masses ($M_\star \sim 10^{8}$--$10^{9}\,M_\odot$), we cross-matched the photometric SDSS data with the DESI-DR1 catalogue, which provides additional spectroscopic redshifts, particularly in regions where SDSS coverage is incomplete. This combined dataset allows us to construct a more statistically robust sample across a wider range of galaxy masses and environments.

Although our sample includes only objects classified as galaxies in SDSS-DR16, contamination by misclassified stars may still occur, especially at faint magnitudes. To reduce that stellar contamination, we applied an additional filtering based on the position of objects in the $m_{r}$--$\langle \mu(R_{50}) \rangle$ plane, where $m_{r}$ is the Petrosian r-band magnitude and $<\mu(R_{50})>$ is the mean surface brightness of the galaxies within their effective radius ($R_{50}$). This plane has been used previously in order to refine the star/galaxy classification \citep[see][]{Sanchez-Janssen2005, Aguerri2020,Zarattini2025}. This filtering is applied to the parent catalogue used for the estimation of the local galaxy overdensity (see Sect.~\ref{subsec:envir}). The resulting SDSS-DR16 + DESI-DR1 galaxy catalogue is formed by a total of 4\,101\,683 objects with redshifts out to $z \sim 0.6$. 

We estimated the stellar mass of the objects in the catalogue by using their $g-r$ colour. In particular, we obtained the stellar mass using the colour-mass relation from \cite{Roediger2015}. 

A more detailed description of the SDSS-DR16+DESI-DR1 catalogue construction, sample selection, photometric and spectroscopic properties, and completeness will be presented in a forthcoming paper (Aguerri et al. 2026, in preparation). That work will provide a comprehensive characterisation of the full galaxy sample and discuss its potential applications for studies of galaxy evolution in the nearby and mid redshift Universe.

\subsection{Selection of compact galaxies} \label{sec:select}

We define our working sample by selecting galaxies with heliocentric recessional velocities $v > 1000$~km\,s$^{-1}$ and redshifts $z < 0.05$. The lower velocity limit ensures a clean extragalactic sample by excluding nearby objects, while the upper redshift limit is chosen to guarantee reliable size measurements of compact galaxies given the spatial resolution of the SDSS imaging.  The typical SDSS seeing, characterised by a full width at half maximum (FWHM) of $\sim 1.3$--$1.4''$, sets the relevant angular resolution scale. We adopt $R_{50} > \mathrm{FWHM}/2$ as our resolution limit in our sample, following previous studies \citep[e.g.,][]{Charbonnier2017, Kim2020, Chen2022}. In physical units, the corresponding physical resolution limit varies with redshift, from approximately 0.14\,kpc at $z\sim0.01$ to approximately 0.6\,kpc at $z\sim0.05$. All galaxies in our working sample satisfy this criterion at their corresponding redshifts.

\begin{figure}
    \centering
    \includegraphics[scale=0.3]{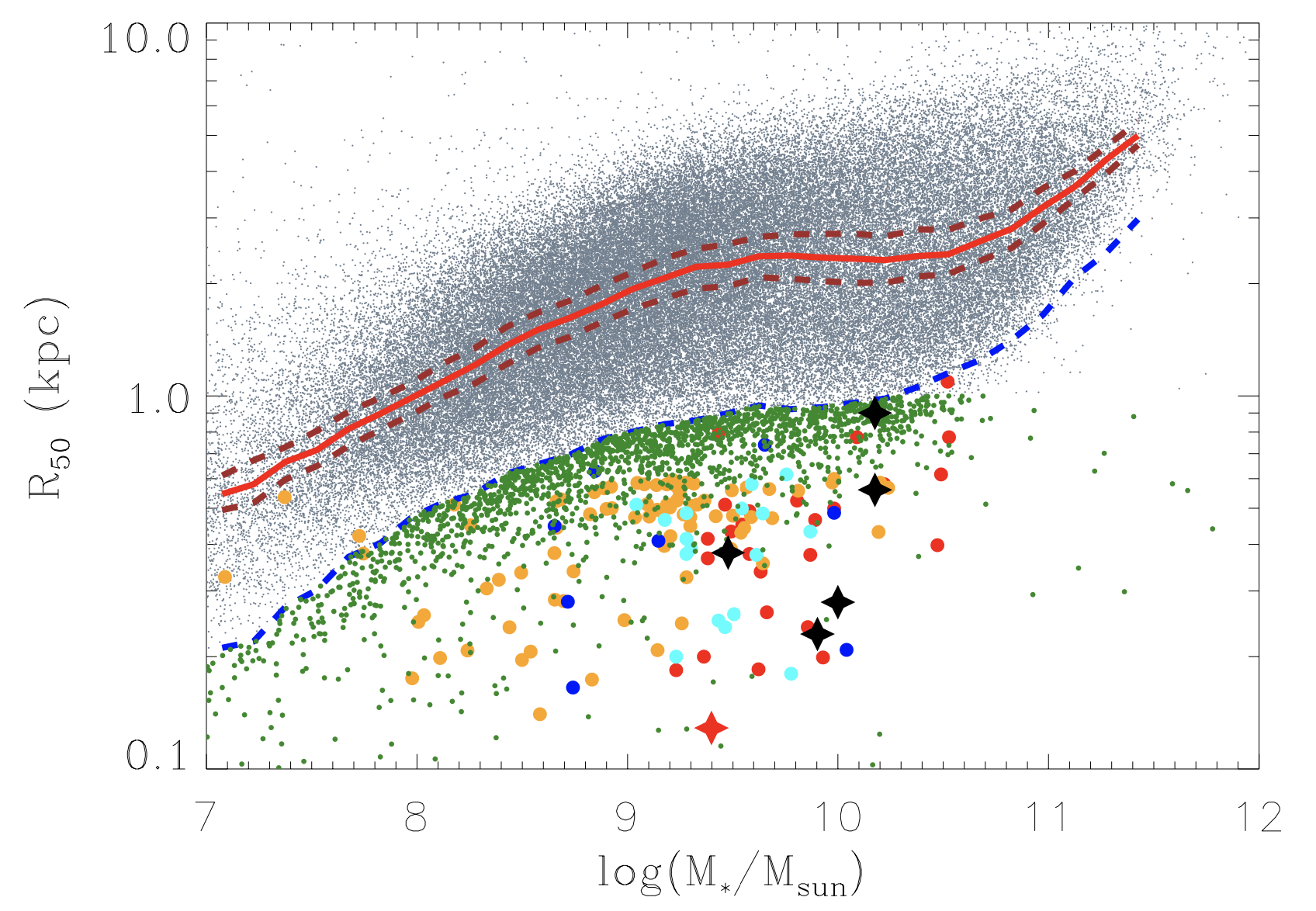}
    \caption{Size--mass relation of galaxies at $z < 0.05$ (gray points). Selected percentiles of the distribution are shown as coloured dashed lines. The median relation (50th percentile) is indicated by the red solid line. Control galaxies are defined between the 40th and 60th percentiles (red dashed lines), while compact galaxies are selected below the 2.275th percentile (blue dashed line) and are highlighted as green points. Colored solid circles represent compact galaxy samples from the literature: blue symbols from \cite{Guerou2015}, cyan symbols from \cite{Janz2016}, red symbols from \cite{Ferre-Mateu2018}, and orange symbols from \cite{Chen2022}.
    Star symbols indicate individual compact galaxies in the nearby Universe, with the red star marking the position of M32.}
    \label{fig:logre_mas}
\end{figure}

Figure \ref{fig:logre_mas} shows the size--mass relation for the 96791 galaxies in our working sample. This relation is consistent with those previously reported in the local Universe using SDSS data \citep[e.g.,][]{Shen2003, Bernardi2014}. We fitted different percentiles of the size--mass distribution to define two distinct galaxy samples. The first sample consists of galaxies lying between the 40th and 60th percentiles. This subset, comprising 18\,763 galaxies, represents systems that follow the average size--mass relation, and we refer to it as the control sample.  The second set of galaxies is selected as those lying below the 2.275th percentile of the size--mass relation, corresponding to objects more than $2\sigma$ below the mean distribution. This compact galaxy sample consists of 1\,824 objects. 

This selection isolates galaxies that are significantly more compact than the average population at fixed stellar mass, allowing us to probe the extreme tail of the size distribution. We refer to this set as the compact galaxy sample. As shown in Fig.~\ref{fig:logre_mas}, our compact sample occupies the same region of the size--mass plane as nearby compact objects such as M32, NGC~4486B, NGC~5846A, NGC~4467, NGC~4478, and CGCG~036-042, as well as previously identified samples of local compact galaxies \cite[][]{Graham2002,Ferrarese2006,Paudel2014,Guerou2015,Janz2016,Ferre-Mateu2018,Ferre-Mateu2021,Chen2022}.  

Although the sample is limited to $z < 0.05$ to ensure reliable size measurements from SDSS photometry, this redshift cut does not guarantee stellar mass completeness across all galaxy types. In particular, the redshifts of low-mass blue galaxies are more easily measured than their red counterparts.  To assess the stellar mass completeness of our sample, we divided the control sample into red and blue galaxies by fitting the red sequence in the colour--mass diagram. The best-fitting relation is given by a slope of 0.087 and an intercept of $-0.148$, with an intrinsic scatter of $\sigma = 0.064$. Blue galaxies are defined as those lying below the fitted red sequence by more than $2\sigma$, while all galaxies above this limit are classified as red. We then examined the distribution of stellar mass as a function of redshift for both populations. 

\begin{figure}
    \centering
    \includegraphics[scale=0.6]{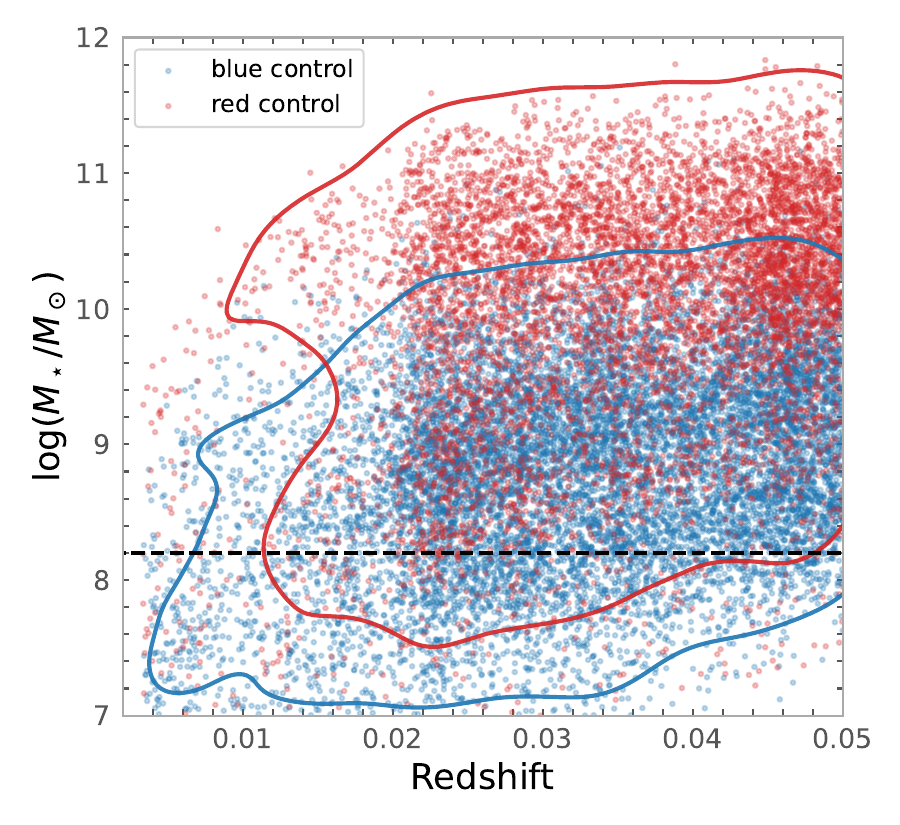}
    \caption{Stellar mass-redshift distribution of blue and red control galaxies. The $2\sigma$ isocontours enclosing the bulk of both populations are overplotted.} 
    \label{fig:mas_redshift}
\end{figure}

Figure~\ref{fig:mas_redshift} shows these distributions, together with the $2\sigma$ isocontours enclosing the bulk of the red and blue galaxy populations. Based on these contours, red galaxies with stellar masses $\log(M_{\star}/M_{\odot}) \gtrsim 8.2$ are detectable across the full redshift range up to $z = 0.05$. We therefore adopt this value as the approximate stellar mass completeness limit of our sample. Applying this stellar mass cut reduces the control and compact galaxy samples to 16\,103 and 1\,525 objects, respectively.

\begin{figure*}
    \centering
    \includegraphics[scale=1.0]{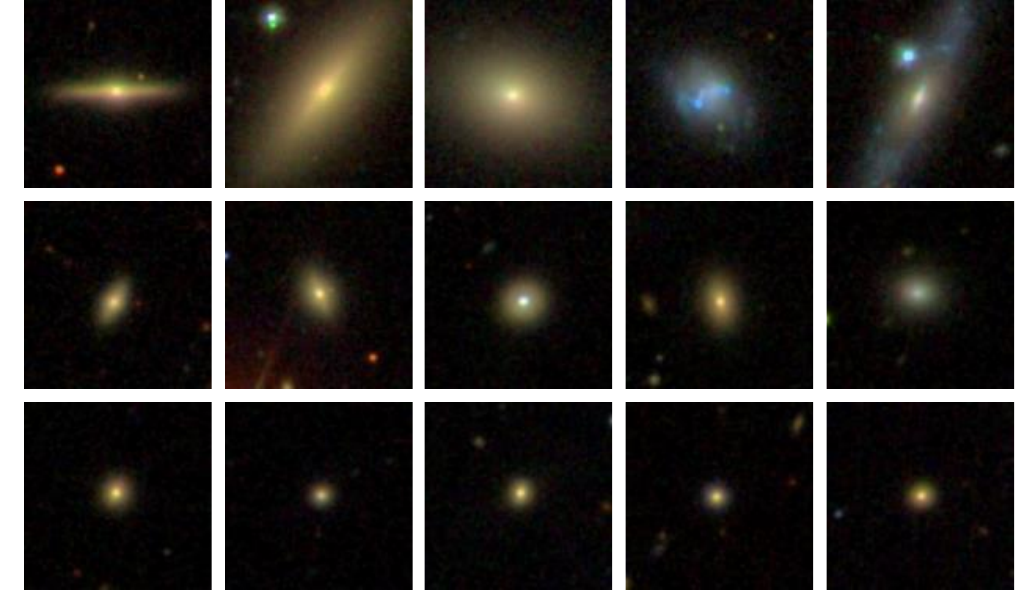}
    \caption{Examples of SDSS colour images illustrating the visual classification of compact galaxy candidates. From top to bottom, the panels show objects flagged as $-1$, $0$, and $1$, respectively. Galaxies with $flag=-1$ correspond to face-on or extended systems with prominent outer structures or compact nuclei, for which the SDSS $R_{50}$ measurements are likely underestimated. Objects with $flag=0$ exhibit a compact core but also display a diffuse or elongated outer component, suggesting the presence of an extended envelope. Galaxies with $flag=1$ are bona fide compact systems, characterised by a compact, centrally concentrated morphology without obvious extended outer structures. Each image has a field of view of $\approx 50''$ on a side.}
    \label{fig:classification}
\end{figure*}

The sample of compact galaxies was further visually examined and flagged into three groups. This visual classification was performed to ensure the reliability of the compact sample and to remove objects whose sizes are affected by photometric or structural biases. The first group consists of objects exhibiting a compact, centrally concentrated morphology, without obvious diffuse envelopes, disc-like components, or other extended outer structures. The second group includes galaxies with a compact morphology but showing a diffuse outer structure, sometimes elongated and resembling a faint disc. The third group comprises extended systems with compact central regions. For these objects, the $R_{50}$ values obtained from the SDSS database are clearly underestimated. The three groups were assigned flags of 1, 0, and $-1$, containing 701, 460, and 364 galaxies, respectively. Figure \ref{fig:classification} shows some illustrative examples of galaxies with the three different flags. Galaxies with $flag = 1$ define our bona fide compact galaxy sample and are used in the remainder of this work.

The final compact galaxy sample remains 701 objects, approximately three times larger than the largest compact galaxy samples previously presented in the literature \citep[e.g.,][]{Chilingarian2015,Kim2020}. These compact systems have effective radii $R_{50} < 1\,\mathrm{kpc}$ and stellar masses $8.2 < \log(M_{\star}/\Msun) < 10.5$, with 304, 356, and 41 galaxies in the stellar mass ranges $8.2 \leq \log(M_{\star}/\Msun) < 9.0$, $9.0 \leq \log(M_{\star}/\Msun) < 10.0$, and $10.0 \leq \log(M_{\star}/\Msun) < 10.5$, respectively. Their structural properties are broadly consistent with those of compact ellipticals, while their stellar masses remain below the regime typically associated with massive compact relic galaxies \citep[e.g.,][]{Trujillo2014,Ferre-Mateu2017,Carr2024}. For consistency, the control sample is also restricted to the same stellar mass range, resulting in 13\,592 galaxies.  Although the stellar mass distributions of the compact and control samples are not identical, additional resampling tests matching the stellar mass distribution of the two samples yield only minor quantitative differences and do not affect our results presented in this work. We therefore retain the original samples in the subsequent analysis.

\subsection{Environmental characterisation} \label{subsec:envir}

To investigate the role of environment in the formation of compact galaxies, we characterise the local density around each object in our compact and control samples. We describe the local environment of the galaxies by their local galaxy density ($\Sigma_{\rm g}$). Similarly to ZA25, we compute $\Sigma_{\rm g}$ by identifying, for each galaxy, its five nearest neighbours within a velocity interval of $\pm 3000$ km s$^{-1}$. 

The spectroscopic galaxy sample from SDSS-DR16 + DESI-DR1 is not complete, implying that galaxy densities derived using only spectroscopic information represent lower limits of the true values. To mitigate this effect, we apply a photometric correction. Specifically, we count galaxies without spectroscopic redshift within the projected area enclosed by the fifth spectroscopic neighbour. We then assume that these galaxies contribute to the local density with the same proportion as those with spectroscopic redshifts, and we include this additional contribution in the final estimate of $\Sigma_{\rm g}$. This correction assumes that galaxies without spectroscopic redshift trace the same underlying density field as those with spectroscopic measurements.

Finally, the local overdensity ($\delta$) is defined as $\delta = (\Sigma_{\rm g} - \langle \Sigma_{\rm g} \rangle)/\langle \Sigma_{\rm g} \rangle$, where $\langle \Sigma_{\rm g} \rangle$ is the mean local galaxy density in the considered redshift bin (see ZA25).

\section{Results} \label{sec:result}

In this section, we present the main results of our analysis of compact galaxies as a function of local environment, in comparison with a control sample representative of ordinary galaxies. We first examine how the relative abundance of compact systems varies across the full range of overdensities, identifying characteristic environmental regimes. We then explore the dependence of these trends on stellar mass, and finally investigate how galaxy properties such as colour, stellar surface mass density, and fraction of red galaxies vary with environment.

\begin{figure*}
    \centering
    \includegraphics[scale=0.48]{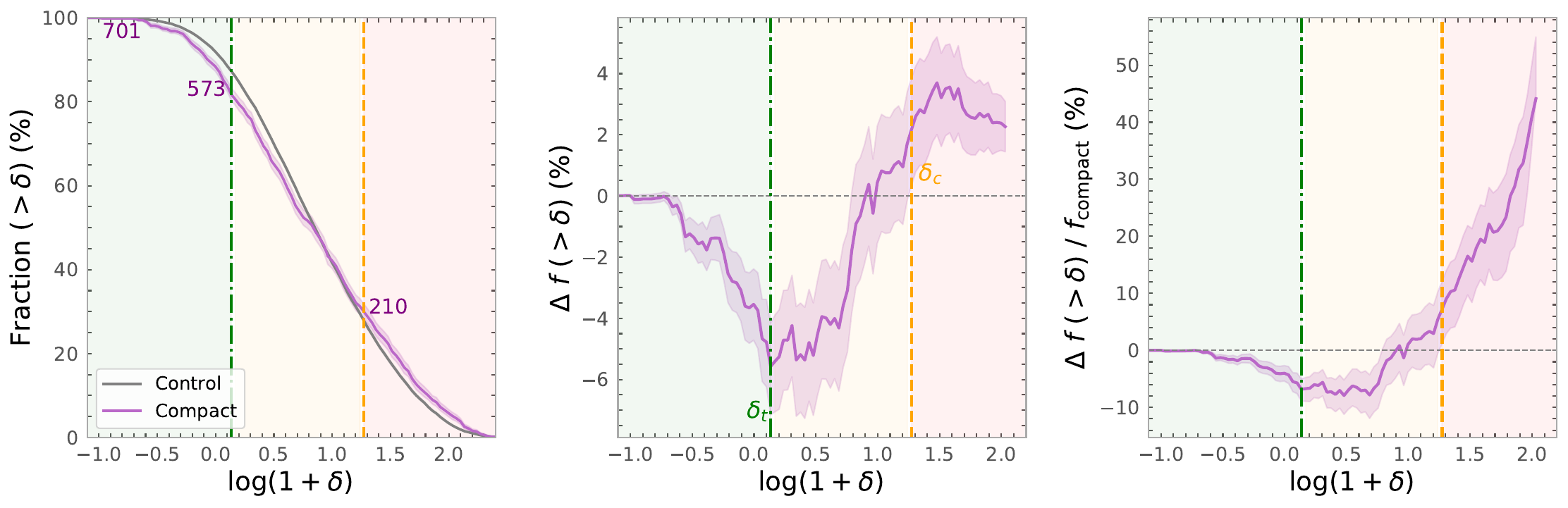} 
    \caption{Environmental dependence of compact galaxies relative to the control sample as a function of $\log(1+\delta)$. 
    \textit{Left:} Complementary cumulative fraction of galaxies residing above a given local overdensity threshold, $f(>\delta)$. 
    \textit{Central:} Difference between compact and control samples, $\Delta f(>\delta) = f_{\rm compact}(>\delta) - f_{\rm control}(>\delta)$. 
    \textit{Right:} Relative excess, defined as $\Delta f(>\delta) / f_{\rm compact}(>\delta)$.
    Solid curves represent the median relations, while the shaded bands around the curves indicate the $16$th--$84$th percentile range derived from $1000$ bootstrap resamplings. 
    The vertical dashed lines mark two characteristic overdensities, the transition overdensity ($\delta_{\rm t}$; green) and the critical overdensity ($\delta_{\rm c}$; orange), which together define three environmental regimes highlighted by the background shading: low-density ($\delta \leq \delta_{\rm t}$), intermediate-density ($\delta_{\rm t} < \delta \leq \delta_{\rm c}$), and high-density ($\delta > \delta_{\rm c}$) environments.
    For the left panel, the numbers indicate the total number of compact galaxies, together with the numbers above $\delta_{\rm t}$ and $\delta_{\rm c}$.
    For the central and right panels, only data with $f_{\rm compact}(>\delta) \geq 5\%$ are shown to ensure statistical robustness.}
    \label{fig:frac}
\end{figure*}

\subsection{A two-threshold environmental transition in the distribution of compact galaxies}

Figure~\ref{fig:frac} presents the environmental dependence of compact galaxies relative to the control sample as a function of local overdensity, quantified by $\log(1+\delta)$. The left panel shows the complementary cumulative fraction of galaxies residing above a given local overdensity threshold, $f(>\delta)$. At low overdensities, the complementary cumulative fraction of control galaxies is higher than that of compact galaxies, indicating that the control sample preferentially inhabits lower-density environments. As the overdensity increases, this trend gradually reverses, and compact galaxies become progressively more dominant at higher-density regions. The Kolmogorov--Smirnov test yields a $p$-value of 0.0031, indicating that the difference between the two distributions is statistically significant, consistent with the systematic trends observed in Fig.~\ref{fig:frac}.

The central panel provides a differential view of this behaviour through $\Delta f(>\delta) = f_{\rm compact}(>\delta) - f_{\rm control}(>\delta)$, where $f_{\rm compact}(>\delta)$ and $f_{\rm control}(>\delta)$ are the complementary cumulative fractions of compact and control galaxies, respectively. This difference highlights how the relative contribution of the two populations varies with overdensity. The function $\Delta f(>\delta)$ reveals three distinct environmental regimes. At low overdensities, $\Delta f(>\delta)$ is negative, indicating that $f_{\rm control}(>\delta) > f_{\rm compact}(>\delta)$, i.e., control galaxies have a higher probability of residing above a given overdensity threshold in this regime. As the overdensity increases, $\Delta f(>\delta)$ becomes increasingly negative until reaching a minimum at a characteristic value that we define as the transition overdensity, $\delta_{\rm t}$. This point corresponds to the maximum difference between the complementary cumulative distributions of the two populations. The transition overdensity, $\delta_{\rm t}$, can be associated with very low-density environments, characteristic of relatively isolated galaxies (see ZA25). Beyond $\delta_{\rm t}$, $\Delta f(>\delta)$ increases with overdensity, reaching a second characteristic value, which we define as the critical overdensity, $\delta_{\rm c}$, where $\Delta f(>\delta)$ becomes statistically positive at the 1$\sigma$ level. At this point, compact galaxies become significantly more abundant relative to the control population above the corresponding density threshold. For overdensities higher than $\delta_{\rm c}$, $\Delta f(>\delta)$ remains significantly positive, indicating that compact galaxies dominate over the control sample at the highest-density environments. 

In the regime below $\delta_{\rm t}$ ($\log (1+\delta) \approx 0.1$), the increasingly negative values of $\Delta f(>\delta)$ indicate that compact galaxies are less likely than control galaxies to be found above a given overdensity threshold. This behaviour is consistent with a population of compact systems preferentially residing in highly isolated environments, where external interactions are rare. 
At overdensities above $\delta_{\rm c}$ ($\log (1+\delta) \approx 1.3$), compact galaxies exhibit higher complementary cumulative fractions than the control sample, indicating that they are preferentially found in the highest-density environments. These overdensities are typically associated with virialised regions of galaxy clusters, and in some cases with their central cores (see ZA25).

The overdensity range between $\delta_{\rm t}$ and $\delta_{\rm c}$ corresponds to intermediate-density environments, such as galaxy groups, the outskirts of clusters beyond the virial radius, and filamentary structures of the cosmic web (see ZA25). In this regime, $\Delta f(>\delta)$ is negative but increases with overdensity, indicating that the relative contribution of control galaxies grows with respect to compact systems. These environments are characterised by moderate galaxy densities and lower velocity dispersions than cluster cores, which favour galaxy-galaxy interactions and mergers. These processes promote structural growth, leading to an increase in galaxy sizes and a corresponding decrease in the fraction of compact systems.

The right panel of Fig.~\ref{fig:frac} shows the normalized difference between the complementary cumulative fractions, defined as $(f_{\rm compact}(>\delta) - f_{\rm control}(>\delta)) / f_{\rm compact}(>\delta)$. This quantity provides a relative measure of the excess or deficit of compact galaxies with respect to the control sample. The overall behaviour closely follows that of $\Delta f(>\delta)$. At low overdensities, the normalized difference is negative, reaching a minimum at $\delta_{\rm t}$, which corresponds to the maximum relative deficit of compact galaxies. As the overdensity increases, the function rises, becoming zero at $\delta_{\rm c}$, where both populations contribute equally. At higher overdensities, the normalized difference becomes increasingly positive, indicating a progressively stronger relative excess of compact galaxies in the densest environments. The amplitude of the normalized difference further highlights the asymmetry between low- and high-density regimes. At overdensities below $\delta_{\rm c}$, the normalized difference always remains above $-10\%$, indicating a relatively mild deficit of compact galaxies with respect to the control sample. In contrast, at overdensities above $\delta_{\rm c}$, the normalized difference rises sharply, reaching maximum values of $\sim 40$--$50\%$. This indicates a strong relative excess of compact galaxies in the highest-density environments, suggesting that such environments may be more effective at enhancing their relative abundance than low-density ones.

\subsection{Mass dependence of the environmental trends} \label{sec:mass}

\begin{figure*}
    \centering
    \includegraphics[scale=0.48]{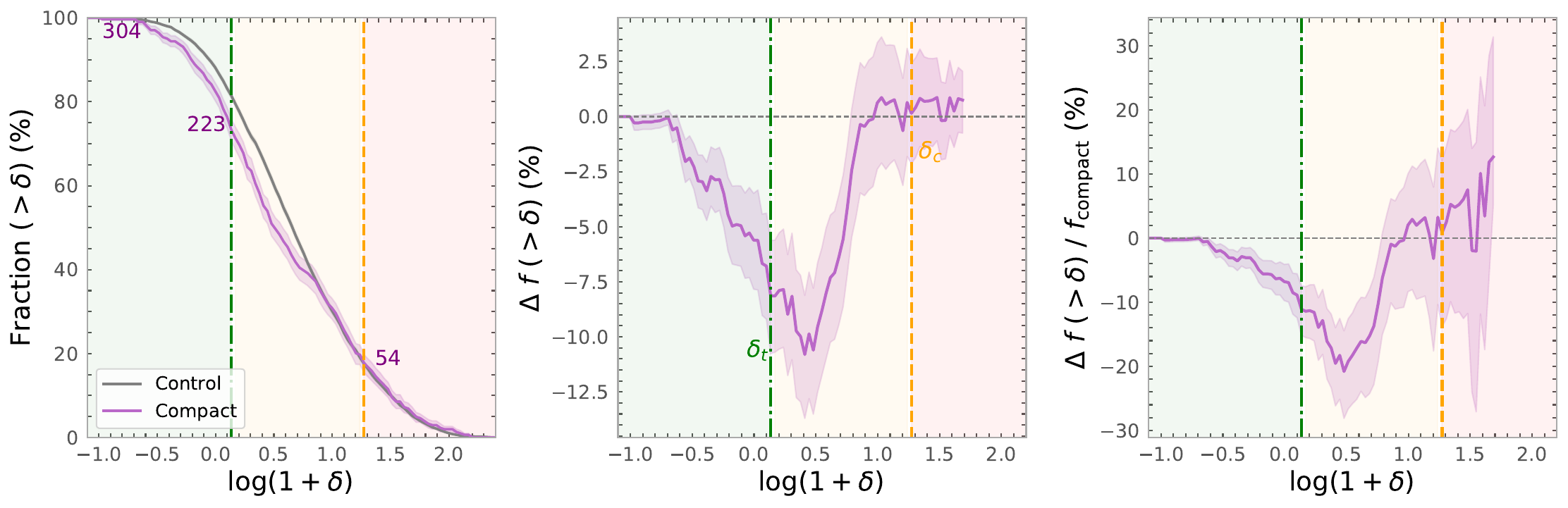}
    \includegraphics[scale=0.48]{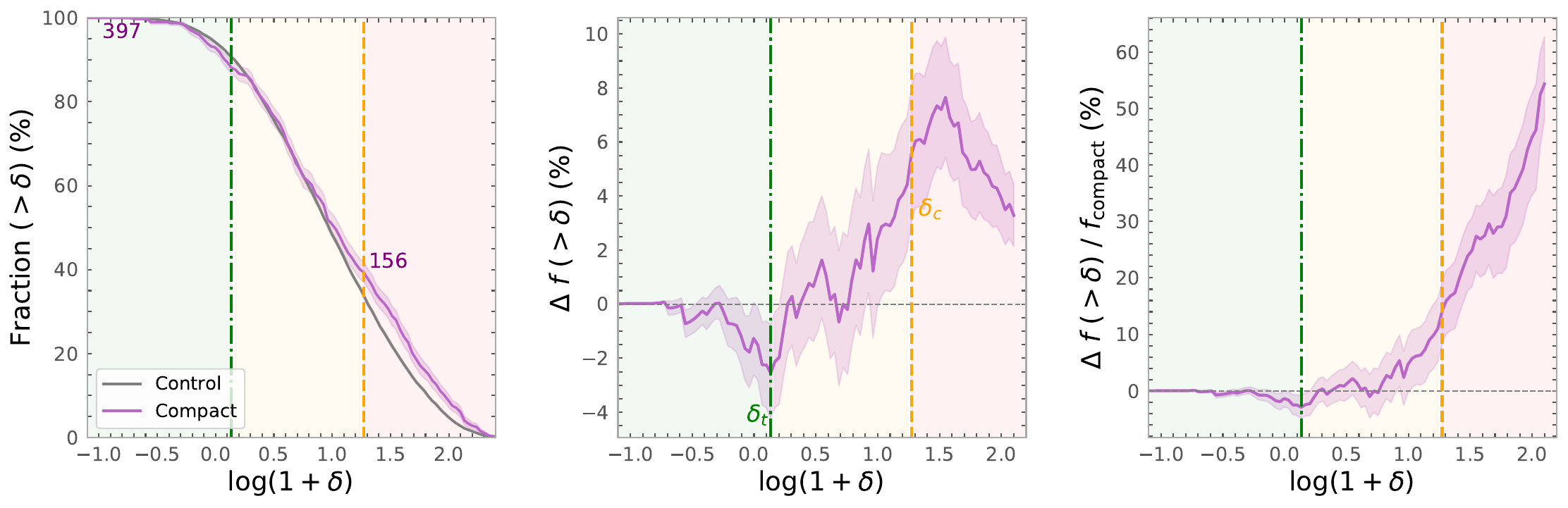}
    \caption{Same as Fig.~\ref{fig:frac}, but split by stellar mass, with top panels showing low-mass galaxies ($8.2 \leq \log M_{\star} < 9.0$) and bottom panels showing high-mass galaxies ($9.0 \leq \log M_{\star} < 10.5$). The vertical lines indicate the transition and critical overdensities ($\delta_{\rm t}$ and $\delta_{\rm c}$) defined from the full compact galaxy sample (see Fig.~\ref{fig:frac}).}
    \label{fig:frac_low_high}
\end{figure*}

We divide the sample into two stellar mass bins, $8.2 \leq \log(M_{\star}/M_{\odot}) < 9.0$ and $9.0 \leq \log(M_{\star}/M_{\odot}) < 10.5$ (see Figs. \ref{fig:frac_low_high}), to explore the mass dependence of the environmental trends. The corresponding subsamples contain $304$ and $397$ galaxies, respectively. The adopted mass boundary at $\log(M_{\star}/M_{\odot}) = 9.0$ roughly corresponds to the transition between classical dwarf and more massive galaxy populations \citep[e.g.,][]{Mateo1998,Geha2012}. In addition, this choice yields subsamples with comparable numbers of objects, allowing for a robust characterisation of the environmental trends. 

The environmental trends exhibit a clear mass dependence with $\delta_{\rm t}$ and $\delta_{\rm c}$ defined from the full compact galaxy sample. Compact dwarfs exhibit a pronounced deficit of systems around $\delta_{\rm t}$ relative to control galaxies, but show little or no significant excess at the highest overdensities ($\delta > \delta_{\rm c}$). This suggests that environmental processes affect low-mass galaxies very efficiently already in intermediate-density environments. Due to their shallow gravitational potentials, dwarf compact galaxies are likely highly vulnerable to tidal disruption, harassment, or accretion by more massive systems, preventing them from surviving in large numbers within the densest cluster regions (see Fig.~\ref{fig:frac_low_high}).

Conversely, more massive galaxies exhibit a much weaker dependence with respect to the control sample at low and intermediate overdensities, with the minimum $\Delta f(>\delta)$ reaching only $\sim -2\%$ around $\delta_{\rm t}$. However, they develop a strong positive excess above $\delta_{\rm c}$. This implies that relatively high-mass compact galaxies survive much more efficiently in dense environments and may preferentially form or evolve there. Their deeper gravitational potentials likely allow them to resist strong environmental effects while simultaneously undergoing stripping of their outer stellar envelopes or environmentally driven structural evolution, ultimately producing compact systems.

These results indicate that the characteristic overdensities identified in the full compact galaxy sample reflect the combined contribution of different mass regimes. Thus, $\delta_{\rm t}$ appears to trace the onset of environmentally driven preprocessing, predominantly affecting low-mass galaxies in intermediate-density environments. On the other hand, $\delta_{\rm c}$ marks the regime where high-mass compact galaxies become preferentially abundant, likely due to enhanced survival probabilities and efficient structural transformation within the densest cluster environments. This scenario is broadly consistent with hierarchical environmental evolution models in which galaxy transformation begins in groups and filaments and culminates in the dense cores of galaxy clusters.

These differences can be interpreted as a consequence of the mass dependence of environmental transformation processes. The intermediate-density regime is expected to be less efficient at driving structural evolution in more massive galaxies, which are already more structurally evolved systems and less susceptible to interactions. In contrast, dwarf galaxies, with their shallower gravitational potentials, are more sensitive to mergers and tidal interactions, allowing compact systems to evolve more efficiently towards extended galaxies in group and filamentary environments. This interpretation is consistent with galaxy downsizing scenarios, in which massive galaxies complete their structural evolution earlier than low-mass systems \citep[e.g.,][]{Cowie1996, Thomas2005, Peng2010}.

\begin{figure*}
    \centering
    \includegraphics[scale=0.48]{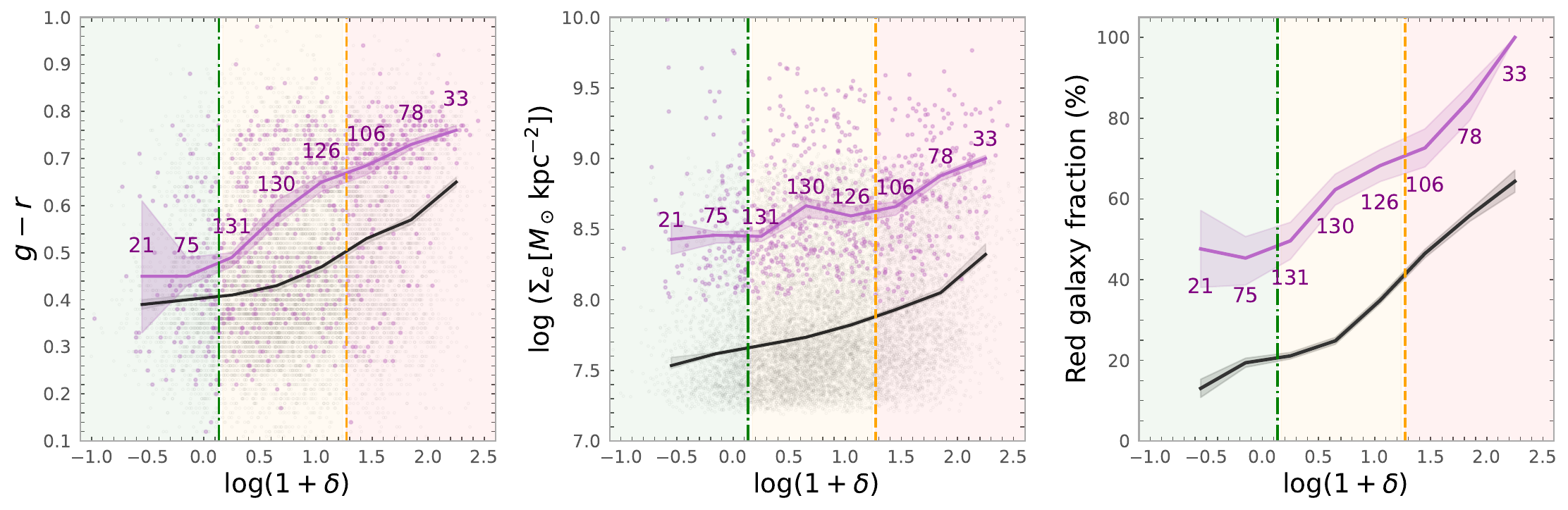}
    \caption{From left to right: environmental dependence of galaxy colour ($g-r$), effective stellar mass surface density ($\Sigma_e = \frac{M_\star}{2\pi R_{50}^2}$), and the red galaxy fraction as a function of local overdensity, $\log(1+\delta)$, for the compact (purple) and control (black) samples. 
    In the left and central panels, points represent individual galaxies, while solid curves show the median trends, with shaded bands around the curves indicating the $16$th--$84$th percentile range derived from $1000$ bootstrap resamplings. 
    In the right panel, solid curves show the fraction of red galaxies within each overdensity bin, with shaded bands indicating the corresponding bootstrap confidence intervals.
    Statistics are computed in fixed overdensity bins requiring at least 15 galaxies; numbers indicate the counts per bin for the compact galaxy sample. Vertical lines mark the two characteristic overdensities ($\delta_{\rm t}$ and $\delta_{\rm c}$), and shaded regions highlight the corresponding environmental regimes.
    }
    \label{fig:prop}
\end{figure*}

\subsection{Galaxy properties as a function of environment}

We analyse the $g-r$ colour of compact and control galaxies as a function of local overdensity (see Fig.~\ref{fig:prop}). Compact galaxies show a clear overall increase in colour with overdensity, transitioning from relatively blue values at low densities ($\langle g-r \rangle \sim 0.45$) to red colours in the highest-density environments ($\langle g-r \rangle \sim 0.8$). Control galaxies exhibit a similar overall trend, evolving from blue ($\langle g-r \rangle \sim 0.38$) to red colours ($\langle g-r \rangle \sim 0.65$) with increasing overdensity. However, two notable differences are observed. First, at fixed overdensity, compact galaxies are systematically redder than control galaxies. Second, the colour evolution of the control sample is relatively weak below $\delta_{\rm t}$, with the median colour remaining nearly constant, and becomes more pronounced at intermediate overdensities between $\delta_{\rm t}$ and $\delta_{\rm c}$. These results indicate that, while both populations follow the general trend of redder colours in denser environments, compact galaxies exhibit an earlier and more continuous transition, whereas control galaxies show a delayed colour evolution with local overdensity.

An interesting feature is that the colour evolution of the control sample appears delayed with respect to the structural transition identified at $\delta_{\rm t}$. While the relative abundance of control galaxies increases above $\delta_{\rm t}$, their median colour remains approximately constant up to intermediate overdensities and only begins to redden significantly between $\delta_{\rm t}$ and $\delta_{\rm c}$. This suggests that structural transformation and quenching are not simultaneous processes. In this scenario, galaxies may first undergo preprocessing in groups and filamentary environments, where mergers and interactions promote structural growth without immediately suppressing star formation. Quenching would then occur at later stages, once environmental processes such as starvation, stripping, or the suppression of gas accretion become more efficient. This interpretation is consistent with evolutionary scenarios in which morphological transformation precedes quenching \citep[e.g.,][]{Peng2010, Wetzel2013, Tacchella2015, Cortese2019}.

Figure \ref{fig:prop} also shows the effective stellar surface mass density ($\Sigma_e = \frac{M_\star}{2\pi R_{50}^2}$) of compact and control galaxies as a function of local overdensity. Both populations exhibit a similar overall trend, with increasing stellar surface density towards higher-density environments. As expected from the sample definition, compact galaxies show systematically higher surface densities than control galaxies by approximately one order of magnitude, independent of environment. The median stellar surface density of control galaxies remains below $\log\Sigma_e = 8.0$ at low and intermediate overdensities, increasing only at the highest densities, where it reaches values of $\log\Sigma_e \sim 8.3$. In contrast, compact galaxies span a narrower range of higher surface densities, with median values between $\log\Sigma_e \sim 8.5$ and $9.0$ across all environments. This indicates that, while both populations become denser in high-density regions, compact galaxies systematically occupy the high end of the surface density distribution. Such high stellar surface densities are commonly associated with spheroid-dominated systems (i.e., $\log\Sigma_e \gtrsim 8.5$), whereas control galaxies are more frequently disc-like objects \citep[e.g.,][]{Kauffmann2003,Kauffmann2006,Erwin2015,Luo2020,Chen2020}.

Moreover, we investigate the red galaxy fraction as a function of local overdensity for both the compact and control samples (the right panel of Fig.~\ref{fig:prop}). The red galaxy fraction, which is commonly used here as a proxy for galaxy quenching, shows a clear and monotonic increase with overdensity in both populations. For compact galaxies, the red fraction rises from $\sim40-45\%$ in low-density environments to $\sim 95\%$ at the highest overdensities. A similar trend is observed for the control sample, although with systematically lower values. At fixed overdensity, compact galaxies exhibit a consistently higher red fraction than control galaxies, with a typical offset of $\sim20-30\%$ across the full overdensity range. This implies that compact systems are more likely to be quenched across all environments compared to control galaxies. Despite this offset, the similar overall trends observed in both samples highlight a strong environmental dependence of quenching. These results indicate that the quenching of both compact and control galaxies is not only modulated by the environment but also regulated by their internal stellar mass distributions. 

\begin{figure*}
    \centering
    \includegraphics[scale=0.48]{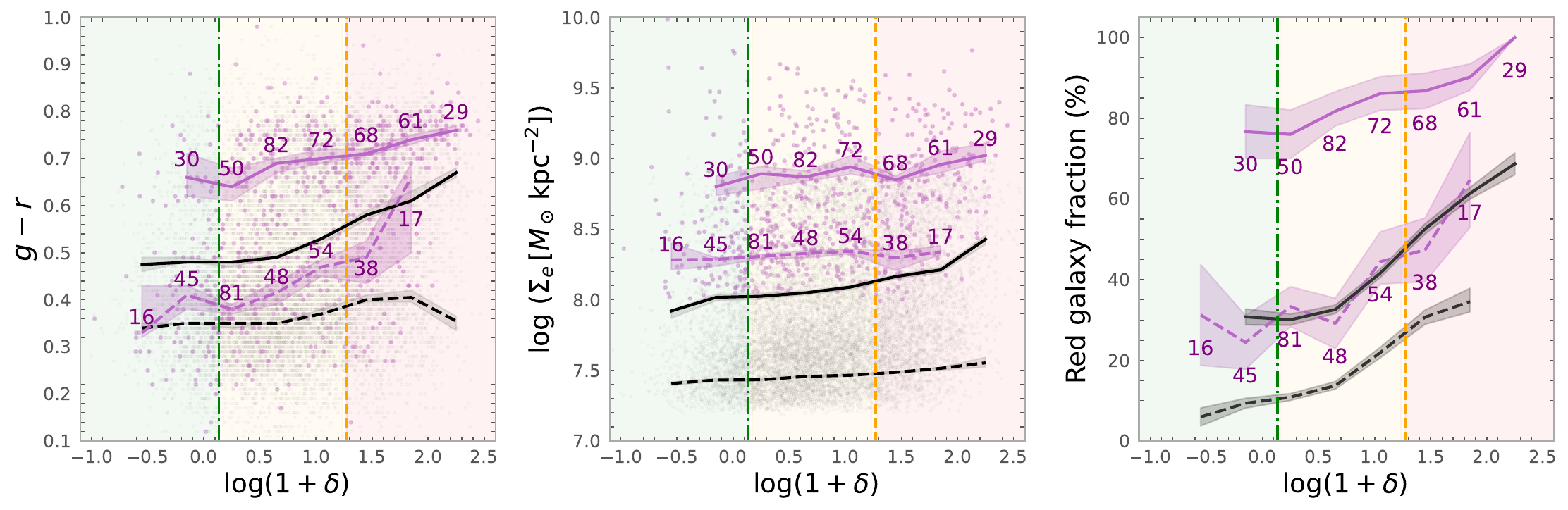}
    \caption{Same as Fig.~\ref{fig:prop}, but separating low-mass galaxies ($8.2 \leq \log M_{\star} < 9.0$; dashed lines) and high-mass galaxies ($9.0 \leq \log M_{\star} < 10.5$; solid lines). Numbers indicate the counts per bin for the compact galaxy sample in the corresponding stellar mass range.}
    \label{fig:prop_mass_split}
\end{figure*}

We further explore the environmental dependence of galaxy properties by separating the compact and control samples into two stellar mass regimes, the low-mass ($8.2 \leq \log M_\star < 9$) and the high-mass ($9 \leq \log M_\star < 10.5$) galaxies (Fig.~\ref{fig:prop_mass_split}). High-mass compact galaxies exhibit uniformly red colours across all environments, with median values increasing only moderately from $\langle g-r \rangle \sim 0.65$ in low-density environments to $\sim0.8$ in the highest-density regions. Their stellar surface mass densities remain extremely high and nearly constant, with $\log\Sigma_e \gtrsim 8.8$ over the full overdensity range. In addition, the red galaxy fraction of high-mass compact galaxies remains very high at all environments, increasing from $\sim75\%$ at low overdensities to almost complete quenching in the densest environments. In contrast, the high-mass control galaxies show systematically bluer colours, lower stellar surface densities, and lower red fractions, together with a stronger environmental dependence. These results indicate that high-mass compact galaxies constitute an already evolved population characterised by old stellar populations, high structural densities, and efficient quenching largely independent of environment.

Low-mass compact galaxies show a markedly different behaviour. Their colours evolve more strongly with overdensity, increasing from relatively blue values in low-density environments to significantly redder colours at high overdensities. Similarly, their red galaxy fractions rise progressively from $\sim30\%$ at low overdensities to $\sim60\%$ in the densest environments. At the same time, their stellar surface mass densities increase only mildly with overdensity and remain systematically lower than those of high-mass compact galaxies. The strongest environmental evolution is observed around the transition overdensity $\delta_t$ (Fig. \ref{fig:prop_mass_split}), where low-mass compact galaxies exhibit the largest deficit relative to the control population (Fig.~\ref{fig:frac_low_high}).

\section{Discussion} \label{sec:discussion}

In this section, we interpret our results within a physical framework connecting galaxy structure, environment, and star formation activity. We first investigate how quenching depends on both compactness and environment, disentangling the relative roles of internal structure and external environmental processes. We then combine these results into a unified evolutionary picture linking the formation and transformation of compact galaxies across different environments.

\subsection{Quenching, structure, and environment} \label{sec:discuss_que}

Understanding how star formation quenching depends on both galaxy structure and environment remains a central question in galaxy evolution \citep[e.g.,][]{Kauffmann2004, Peng2010, Woo2015, Wang2018, Bluck2020}. In the previous section, we showed that the abundance of compact galaxies exhibits two characteristic environmental transitions associated with changes in the relative fraction of compact and control galaxies across different overdensity regimes. The observed trends allow us to explore how quenching relates to galaxy structure and environment by comparing the behaviour of galaxies as a function of stellar surface mass density and local overdensity.

The environmental dependence on galaxy quenching becomes evident when the red galaxy fraction is considered as a function of local overdensity (the right panel of Fig.~\ref{fig:prop}). For both compact and control galaxies, the red fraction increases monotonically with overdensity, implying that denser environments are associated with a higher probability of quenching \citep[see][]{Hogg2003}. At fixed overdensity, compact galaxies exhibit systematically higher red fractions than control galaxies, consistent with their higher surface densities. At the same time, the persistence of a significant offset between compact and control galaxies at fixed overdensity indicates that environment alone cannot explain the observed differences in quenching between the two populations.

Since this offset is present across the full range of overdensities, the enhanced red galaxy fraction of compact galaxies is likely related to their internal structure, namely their higher stellar mass densities and more centrally concentrated mass distributions. Although stellar mass correlates with overdensity in both samples, the lack of significant mass differences at fixed overdensity suggests that stellar mass alone is unlikely to fully explain the observed offset in red galaxy fraction. This supports the interpretation that the mass distribution (compactness) plays a key role in the observed quenching differences between the two samples. In this picture, environment modulates the overall efficiency of quenching, increasing the red galaxy fraction towards higher overdensities in both populations, while compactness is closely associated with the baseline probability that a galaxy is quenched. This suggests that the distribution of stellar mass within galaxies may be more directly linked to quenching than environment alone \citep[][]{Franx2008, Fang2013, Woo2015, Bluck2020, Hong2023}.

The stellar mass dependence of the red galaxy fraction further reveals significant differences in the quenching behaviour of compact galaxies. Relatively high-mass compact galaxies exhibit consistently high red galaxy fractions ($\gtrsim70\%$) across the full overdensity range, together with extremely high stellar surface mass densities and red colours. Their weak environmental dependence suggests that quenching in these systems is primarily regulated by internal structural properties, rather than by ongoing environmental transformation.

In contrast, low-mass compact galaxies show a much stronger environmental dependence. Their red galaxy fractions increase progressively with overdensity, rising from $\sim30\%$ in low-density environments to $\sim60\%$ at the highest overdensities. This transition occurs primarily between the transition overdensity $\delta_t$ and the critical overdensity $\delta_c$, coincident with the environmental regime where low-mass compact galaxies become relatively deficient with respect to the control population. This behaviour indicates that low-mass compact galaxies experience a strong environmental modulation of their quenching in groups, filaments, and cluster infall regions.

This interpretation naturally connects with the framework in which quenching is governed by both internal and external processes. Structural quenching models propose that the buildup of a dense stellar component can stabilize the gas and suppress star formation \citep[e.g.,][]{Martig2009, Genzel2014, Tacchella2015}, providing a physical mechanism linking $\Sigma_e$ and quenching. In this context, high-mass compact galaxies may be structurally predisposed to quench. However, the environmental dependence observed in the low-mass compact objects indicates that such structural conditions are not sufficient on their own, and that external processes play a key role in triggering or accelerating quenching. Mechanisms such as ram-pressure stripping, tidal interactions, and the suppression of gas accretion are expected to become increasingly important in dense environments \citep[e.g.,][]{Boselli2006, Peng2010, Peng2012, Wetzel2013}.

An alternative, but not mutually exclusive, interpretation is that the observed correlation between $\Sigma_e$ and quenching does not reflect a direct causal connection, but instead arises from the evolutionary history of galaxies. In particular, \citet{Lilly2016} showed that a sharp apparent surface density threshold can emerge naturally from the combined effects of galaxy size evolution and progenitor bias, even in models where quenching depends primarily on stellar mass. In this scenario, quenched galaxies are denser because they formed earlier, rather than because high density directly suppresses star formation. This interpretation is also broadly consistent with models in which relatively high-mass compact galaxies represent systems that formed the bulk of their stellar mass at earlier epochs and subsequently evolved with limited structural evolution \citep[e.g.,][]{vanDokkum2010, Wellons2016, Wang2018b}.

Taken together, the observed mass dependence indicates that the relative importance of structure and environment in regulating quenching strongly depends on stellar mass. In high-mass compact galaxies, quenching appears primarily linked to internal structural properties, whereas in low-mass compact galaxies, environmental effects play a much stronger role. These results highlight that the connection between compactness and quenching cannot be understood independently of stellar mass and environment.

\subsection{A unified vision for compact galaxy formation}

The results presented in this work support the scenario that compact galaxies do not constitute a single homogeneous population formed through a unique evolutionary pathway, but instead arise through multiple channels linked to stellar mass and environment \citep[e.g.,][]{Ferre-Mateu2021,GrebolTomas2023,Carr2024}.

The environmental distributions shown in Figs.~\ref{fig:frac} and \ref{fig:frac_low_high} reveal the existence of two characteristic overdensity regimes, defined by the transition overdensity $\delta_t$ and the critical overdensity $\delta_c$. Compact galaxies are relatively more abundant than the control population at low overdensities, become relatively deficient in intermediate-density environments around $\delta_t$, and exhibit a significant excess at the highest overdensities above $\delta_c$. However, this behaviour strongly depends on stellar mass. Low-mass compact galaxies are responsible for the deficit observed around $\delta_t$, while the excess at $\delta>\delta_c$ is almost entirely driven by high-mass compact galaxies.

This environmental segregation is closely connected with the stellar population and structural trends presented in Figs.~\ref{fig:prop} and \ref{fig:prop_mass_split}. Compact galaxies systematically exhibit redder colours, higher stellar surface mass densities, and larger red galaxy fractions than the control sample at nearly all environments. Nevertheless, the environmental dependence of these properties differs substantially between low- and high-mass compact galaxies.

High-mass compact galaxies display uniformly red colours, extremely high stellar surface mass densities, and red galaxy fractions exceeding $\sim70\%$ even in relatively low-density environments. Furthermore, they show a strong excess in the densest environments above $\delta_c$. This suggests that relatively high-mass compact galaxies are unlikely to be undergoing strong environmental transformation, but may instead represent an evolved population capable of surviving in the densest environments of the local Universe. Their properties are broadly consistent with scenarios proposed for compact relic systems such as NGC~1277 \citep[e.g.,][]{Trujillo2014,Ferre-Mateu2017} and FS90~192 \citep[][]{Caso2024}, which are interpreted as local survivors or analogues of the high-redshift ``red nugget'' population. In this picture, compact galaxies may have formed early through highly dissipative processes at high redshift and experienced rapid quenching shortly afterwards. Their compact structure and high stellar densities make them highly resistant to tidal disruption, while the large velocity dispersions of cluster environments suppress major mergers and therefore limit subsequent size growth. As a consequence, dense environments may act not only as sites where massive compact galaxies survive preferentially, but also as environments that preserve relic systems by preventing their later structural evolution.

In contrast, low-mass compact galaxies show a much stronger dependence on environment. Their colours, stellar surface mass densities, and red galaxy fractions evolve progressively with overdensity, particularly between $\delta_t$ and $\delta_c$. At the same time, low-mass compact galaxies do not exhibit a significant excess above the critical overdensity $\delta_c$. This behaviour suggests that low-mass compact galaxies are closely linked to environmentally driven transformation processes acting in groups, filaments, and cluster infall regions. Their increasing red galaxy fractions and structural evolution towards higher stellar surface densities are consistent with preprocessing mechanisms such as tidal stripping, harassment, ram-pressure stripping, and gas depletion. In this scenario, low-mass compact galaxies may represent transitional systems undergoing environmentally driven compaction and quenching before entering the densest cluster regions.

The absence of a strong excess of low-mass compact galaxies at the highest overdensities further suggests that these systems do not survive efficiently within cluster cores. Their shallow gravitational potentials likely make them highly vulnerable to tidal disruption and dynamical destruction. Some systems may be transformed into ultra-compact dwarfs, accreted by central galaxies, or contribute to the intracluster light. Therefore, the densest cluster environments may act as efficient filters where the destruction timescale of compact dwarf galaxies becomes comparable to or shorter than their formation timescale.

Taken together, our results support a unified evolutionary framework in which compact galaxy formation and survival depend strongly on stellar mass. Relatively high-mass compact galaxies appear to be predominantly relic-like systems formed at early epochs and preserved within dense environments, whereas low-mass compact galaxies are more likely transient products of ongoing environmentally driven transformation processes. In this picture, the transition overdensity $\delta_t$ marks the onset of efficient preprocessing acting primarily on low-mass galaxies, while the critical overdensity $\delta_c$ corresponds to the regime where only the densest and most structurally resilient compact systems survive efficiently.

This scenario naturally explains the coexistence of multiple compact galaxy populations in the local Universe and reconciles several apparently contradictory formation mechanisms proposed in previous studies. Rather than representing mutually exclusive channels, relic formation, tidal stripping, ram-pressure-induced compactification, and environmental preprocessing may all contribute to the compact galaxy population, but with their relative importance depending strongly on galaxy mass and environment.

\section{Conclusions} \label{sec:conclusions}

We have constructed the largest sample of compact galaxies in the nearby Universe ($z<0.05$), comprising 701 systems with $8.2 \lesssim \log(M_{\star}/M_{\odot}) \lesssim 10.5$, and analysed their environmental dependence using a homogeneous and continuous characterisation of local overdensity.

Our results reveal that the environmental dependence of compact galaxies is characterised by two main overdensity transitions, defined by the transition overdensity $\delta_{\rm t}$ and the critical overdensity $\delta_{\rm c}$. Below $\delta_{\rm t}$, compact galaxies are relatively more abundant than the control population, whereas between $\delta_{\rm t}$ and $\delta_{\rm c}$ their relative abundance decreases significantly. Above $\delta_{\rm c}$, compact galaxies become increasingly overabundant again, although this behaviour strongly depends on stellar mass.

The stellar mass dependence reveals two distinct behaviours within the compact galaxy population. Low-mass compact galaxies are strongly affected by environment, exhibiting a pronounced deficit between $\delta_{\rm t}$ and $\delta_{\rm c}$ together with a progressive increase in the red galaxy fraction and redder colours towards higher overdensities. This suggests that low-mass compact galaxies experience strong environmental processing in groups, filaments, and cluster infall regions, where mechanisms such as tidal interactions, ram-pressure stripping, harassment, and starvation likely play an important role in driving their structural and star formation evolution. However, low-mass compact galaxies do not exhibit a significant excess above the critical overdensity $\delta_{\rm c}$, suggesting that they may not survive efficiently within the densest cluster environments.

In contrast, high-mass compact galaxies show weak environmental dependence at intermediate overdensities, but dominate the excess population above $\delta_{\rm c}$. These systems exhibit uniformly red colours, high stellar surface mass densities, and high red galaxy fractions across the full overdensity range. Their properties are consistent with a population of compact systems that are capable of surviving in dense environments, analogous to local compact relic galaxies such as NGC~1277.

The strong connection between stellar surface mass density and red galaxy fraction suggests that internal structure plays a key role in regulating star formation activity. However, the stronger environmental dependence observed in low-mass compact galaxies indicates that external environmental processes become increasingly important at low stellar masses, while high-mass compact systems appear to be primarily regulated by their internal structural properties. Our results therefore suggest that the relative importance of internal structure and environment in driving quenching depends strongly on stellar mass.

Overall, our results support a unified evolutionary framework in which compact galaxies do not represent a single homogeneous population, but instead arise from multiple evolutionary channels whose relative importance depends on both stellar mass and environment. In this picture, massive compact galaxies are likely associated with relic-like systems that survive efficiently in dense environments, whereas low-mass compact galaxies are more closely linked to environmentally driven transformation processes acting in groups, filaments, and cluster infall regions.

Future studies based on detailed stellar populations, star formation histories, and spatially resolved spectroscopy of compact galaxies across different environmental regimes will be essential to disentangle the relative importance of the different evolutionary pathways and to distinguish between relic systems and environmentally transformed compact galaxies.

\begin{acknowledgements}
GC, JALA and CdB acknowledge support from the Agencia Estatal de Investigación del Ministerio de Ciencia, Innovación y Universidades (MCIU/AEI) under  grant WEAVE: EXPLORING THE COSMIC ORIGINAL SYMPHONY, FROM STARS TO GALAXY CLUSTERS and the European Regional Development Fund (ERDF) with reference PID2023-153342NBI00/10.13039/501100011033. CdB also acknowledges support from a Beatriz Galindo Senior Fellowship (BG22/00166) from the MICIU, and The University of La Laguna (ULL) and the Department of Economy, Knowledge, and Employment of the Government of the Canary Islands (2024/347). SZ acknowledges the financial support provided by the Governments of Spain and Arag\'on through their general budgets and the Fondo de Inversiones de Teruel, the Aragonese Government through the Research Group E16\_23R, and the Spanish Ministry of Science and Innovation and the European Union - NextGenerationEU through the Recovery and Resilience Facility project ICTS-MRR-2021-03-CEFCA. Funding for the Sloan Digital Sky Survey IV has been provided by the Alfred P. Sloan Foundation, the U.S. Department of Energy Office of Science, and the Participating Institutions. SDSS acknowledges support and resources from the Center for High-Performance Computing at the University of Utah. The SDSS web site is www.sdss4.org. SDSS is managed by the Astrophysical Research Consortium for the Participating Institutions of the SDSS Collaboration including the Brazilian Participation Group, the Carnegie Institution for Science, Carnegie Mellon University, Center for Astrophysics Harvard \& Smithsonian (CfA), the Chilean Participation Group, the French Participation Group, Instituto de Astrofísica de Canarias, The Johns Hopkins University, Kavli Institute for the Physics and Mathematics of the Universe (IPMU) / University of Tokyo, the Korean Participation Group, Lawrence Berkeley National Laboratory, Leibniz Institut für Astrophysik Potsdam (AIP), Max-Planck-Institut für Astronomie (MPIA Heidelberg), Max-Planck-Institut für Astrophysik (MPA Garching), Max-Planck-Institut für Extraterrestrische Physik (MPE), National Astronomical Observatories of China, New Mexico State University, New York University, University of Notre Dame, Observatório Nacional / MCTI, The Ohio State University, Pennsylvania State University, Shanghai Astronomical Observatory, United Kingdom Participation Group, Universidad Nacional Autónoma de México, University of Arizona, University of Colorado Boulder, University of Oxford, University of Portsmouth, University of Utah, University of Virginia, University of Washington, University of Wisconsin, Vanderbilt University, and Yale University. 
\end{acknowledgements}

\bibliography{bibliografia}

\end{document}